\documentstyle[12pt,iopconf]{article}

\def \be{\begin{equation}}
\def \ee{\end{equation}}
\def \bes{\begin{eqnarray}}
\def \ees{\end{eqnarray}}

\begin{document}
\title{Causality in spin foam models for quantum gravity\footnote{This
work is a much shortened presentation of the results published in
\cite{causal}, to which we refer for a more detailed account and a complete set of references.}}
\author{Etera R Livine\dag
and Daniele Oriti\ddag}
\affil{\dag\ Centre de Physique Th{\'e}orique,
Campus de Luminy,  \\Case 907,
13288 Marseille cedex 9, France; livine@cpt.univ-mrs.fr}
\affil{\ddag\ D.A.M.T.P. - C.M.S. - University of Cambridge \\
Wilberforce Road, Cambridge CB3 0WA, UK; d.oriti@damtp.cam.ac.uk}
\beginabstract
We describe how the Barrett-Crane spin foam model defines transition
amplitudes
for quantum gravity states and how causality can be consistently
implemented in it.
\endabstract
\section{Introduction}
Spin foam models \cite{Baez}\cite{danrev}
have emerged recently as a new promising approach to the
construction of a quantum theory of gravity, being an explicit
implementation of the path integral approach to quantum gravity,
where amplitudes for gravity states are defined as a sum over all the
4-geometries interpolating between given boundary 3-geometries,
weighted by the action for gravity, with an additional sum over 4-manifolds: 
\begin{equation}
Z(h_1,h_2)\,=\,\sum_{\mathcal{M}}\int_{h_1,h_2}{\mathcal{D}}g\, \e^{i\,S_{gr}}.
\end{equation}
Spin foam models are constructed out of only
combinatorial and algebraic data (coming from the representation
theory of the Lorentz group) and continuum geometric
structures are expected to emerge as
an approximation in some appropriate limit.
Causality is of course a crucial ingredient for this to happen,
since a classical metric is determined almost
completely  by the
causal structure of spacetime, with the
remaining degree of freedom being just a length scale. It is also
a crucial element to understand
what kind of transition amplitudes spin foam models define.
In fact, many different amplitudes may be given by a path
integral realization even for the simple case of a relativistic
particle in flat space, with action
$S(x)\,=\,\int_{\lambda_1}^{\lambda_2}( p_\mu
\dot{x}^\mu\,-\,T\,\mathcal{H}) \d \lambda $, where
${\mathcal{H}}=p_{\mu} p^{\mu} + m^{2} = 0 $ is the Hamiltonian constraint
 and $T$ is the proper
time elapsed between the initial and final state. An integral
over both $T<0$ and $T>0$ yields the Hadamard Green function:
\bes
G_H(x_1,x_2)&=&\langle x_2\mid x_1\rangle =
\int_{-\infty}^{+\infty} \d T
\int \d x \d p\, e^{i\,\int \d\lambda\,(p x - T
\mathcal{H})}=
\int \d x\, \d p\,\delta(p^2 +
m^2)\,e^{i\int \d\lambda\,\, x\, p}= \nonumber \\  &=& \,
G^+(x_1,x_2)\,+\,G^-(x_1,x_2)
\,=\,G^+(x_1,x_2)\,+\,G^{+}(x_2,x_1)\,=\,G_H(x_2,x_1), \label{eq:had} \ees where the
$G^{\pm}$ are the Wightman functions. $G_H$ solves the Klein-Gordon equation in both its
arguments, does not register any order between them, is an a-causal
amplitude between physical states and defines a generalized projector operator from
kinematical states onto solutions of the Hamiltonian constraint:
$G_H(x_1,x_2)=\langle x_2\mid x_1\rangle_{phys}=_{kin}\langle x_2 \mid
{\mathcal{P}}_{{\mathcal{H}}= 0}\mid x _{1}\rangle_{{\it kin}}$ .
An integral over only $T>0$, so that $\mid x_2\rangle$ lies always in the future
of $\mid x_1\rangle$, gives the Feynman propagator or causal amplitude:
\be
G_F(x_1,x_2)=\langle x_2\mid x_1\rangle_C\,=\,\theta(x_1^0-x_2^0)\,G^+(x_1,x_2)+
\theta(x_2^0-x_1^0)\,G^-(x_1,x_2),
\ee
which is not a solution of the Klein-Gordon equation, does not realize the projection operator, but is a transition
amplitude which takes into account causality.
Therefore, after having recognised a spin foam model
as a realization of the a path integral for
quantum gravity, one has still answer several key questions:
what kind of amplitude does it define? is it an implementation of
the projector operator or a realization of the Feynman propagator?
if it defines a projector, where is encoded the ${\rm \bf
Z}_2$ symmetry between $T$ and $-T$? how to break
such a symmetry and implement causality? 

\section{The (quantum) geometry of the Lorentzian Barrett-Crane model}
The Barrett-Crane spin foam model\cite{bc2}\cite{danrev}\cite{Baez} is a path integral quantization of the action: \be
S(\omega,B,\phi)=\int_{\mathcal{M}}\left[ B^{IJ}\wedge
F_{IJ}(A)-\frac{1}{2}\phi_{IJKL}B^{IJ}\wedge B^{KL}\right]
\ee
which is a $BF$ theory with variables
a 2-form $B^{IJ}_{\mu\nu}$ and a 1-form connection $A^{IJ}_{\mu}$ (with
curvature $F^{IJ}_{\mu\nu}$), all
with values in $so(3,1)$,
but with a constraint on the $B$ field enforced
by the Lagrange multiplier
$\phi_{IJKL}$. The constraints on the $B$ field have
four sectors of solutions\cite{danrev}:
$B^{IJ}=\pm e^{I}\wedge e^{J}$ and 
$B^{IJ}=\pm\frac{1}{2}\epsilon^{IJ}\,_{KL}e^{K}\wedge e^{L}$,
so in one of these sectors: $S\rightarrow S_{EH}=\int
\epsilon_{IJKL}e^{K}\wedge
e^{L}\wedge F^{IJ}$, i.e. the theory reduces to pure 1st order Einstein gravity.
The other sector, differing by a global change of sign only,
is classically equivalent to this, while the other two have
no geometrical interpretation. The ${\rm \bf Z}_2$ symmetry between the two geometric
sectors directly affects the path integral quantization, since a change
of sign in the $B$ field is equivalent to a change of sign in the
lapse function, and thus in the proper time.
We replace the
continuum manifold by a simplicial complex, or by its dual 2-complex, and the
continuum fields by discrete variables. In particular, the $B$ field is associated to the
triangles of the triangulation by: $B^{IJ}(t)=\int_t
B^{IJ}_{\mu\nu}(x) dx^\mu \wedge dx^\nu\in so(3,1)\simeq \wedge^2({\rm
\bf R}^{3,1})$, and the constraints on the $B$ field become
constraints on the bivectors
$B^{IJ}$
\cite{bc2}.
The quantization\cite{bc2} then proceeds by associating to each
triangle an irreducible representation of the Lorentz group in the
principal unitary series (labelled by $n\in{\rm
\bf N}$ and $\rho\in{\rm \bf R}_+$), with the identification:
$B^{IJ}(t)\leftrightarrow \ast J^{IJ}((n,\rho)_t)$,
where the $J$'s are the generators of the Lorentz algebra, and
assigning to each tetrahedron a tensor in the
tensor product of the four representation spaces of its triangles. The
constraints on the bivectors then become constraints on the representations used and on these
tensors.
To obtain the Barrett-Crane partition function one may proceed in
several ways \cite{danrev}\cite{boundary}\cite{alejandro1}. In each case one starts
from $BF$ theory and imposes the BC
constraints on this using suitable projectors. 
The resulting model, using only the simple representations $(0,\rho)$, 
is:
\be
Z\,=\,(\prod_f
\int_{\rho_f}d\rho_f\,\rho_f^2)\,(\prod_{v,
e_v}\int_{H^+}dx_{e_v})\,\prod_e
{\mathcal{A}}_e(\rho_k)\,\prod_v{\mathcal{A}}_v (\rho_k, x_i)
\label{eq:Z}
\ee
with the amplitudes for vertices (4-simplices) being given by:
\bes
{\mathcal{A}}_v(\rho_k,
x_i)=K^{\rho_1}(x_1,x_2)K^{\rho_2}(x_2,x_3)K^{\rho_3}(x_3,x_4)K^{\rho_4}(x_4,x_5)K^{\rho_5}(x_1,x_5)
\nonumber \\ K^{\rho_6}(x_1,x_4)
K^{\rho_7}(x_1,x_3)K^{\rho_8}(x_3,x_5)K^{\rho_9}(x_2,x_4)
K^{\rho_{10}}(x_2,x_5) \label{eq:ampl},
\ees
and those for edges (tetrahedra) being also explicitely known and
considered part of the measure\cite{causal}.
The functions $K$ (one for each face (triangle)) have the explicit
expression:
$K^{\rho_k}(x_i,x_j)=\frac{2\sin(\eta_{ij}\rho_k/2)}{\rho_k\sinh\eta_{ij}}$,
where $\eta_{ij}=\cosh^{-1}(x_i\cdot x_j)$  is the hyperbolic distance
between the points $x_i$ and $x_j$ on the hyperboloid
$H^+=\{x^{\mu}\in{\rm \bf R}^{3,1}/x\cdot x=1, x^0>0\}$.
The partition function above should be understood
as just a term within a sum over triangulations or over
2-complexes, to restore the full dynamical content of the quantum gravity theory. Upon
quantization, the triangle areas are given by
$A_t^2=B_t^{IJ}B_{tIJ}=-J^{IJ}(\rho_t)J_{IJ}(\rho_t)=\rho_t^2+1>0$, so
the $\rho$'s determine the areas of the triangles, and we see that all
the triangles are spacelike (consequently, also all the tetrahedra are spacelike). The same result can also be confirmed by
a canonical analysis of the area operator\cite{link}. The $x\in
H^{+}$ variables are un-oriented normals to the tetrahedra of the
manifold, the oriented normals being $n_i=\alpha_i
x_i$, with $\alpha=\pm 1$ 
for future- and $\alpha=-1$ for past-oriented $n$. 
The Barrett-Crane model corresponds
to a piecewise flat
manifold, with patches of flat space-time, the 4-simplices, glued
together along their common tetrahedra. To each 4-simplex is attached a local reference
frame, and there is a non-trivial
connection rotating from one to another. The Lorentz invariance at 
each
4-simplex may be used to fix one of the
vectors $x$, so the true
variables are
the hyperbolic distances $\eta$ between any two of them. These in
turn correspond to the dihedral angles, $\theta_{ij}^{\sigma}=\alpha_i\alpha_j\eta_{ij}$, between two tetrahedra sharing a triangle. Therefore the underlying classical theory for
the Barrett-Crane model, somehow hidden in the above formulation, is a first order formulation of Regge Calculus
based on angles and areas as fundamental variables\cite{causal}, with the angles constrained by the Schl{\"a}fli
identity:  
$\sum_{i\neq j}\,A_{ij}\,\d\theta_{ij}\,=\,0$.
What are the quantum gravity states in the Barrett-Crane model? Let us
consider a spin foam with boundary, this being made of 4-valent vertices glued to form an oriented graph. 
The boundary states are Lorentz invariant
functionals of group elements $g_e$ living on the edges of the graph
and normals $x_v$ at each vertex,
invariant with respect to $SU(2)$ at each vertex and
edge (simplicity
constraints). This space of functionals
is endowed with the
$SL(2,{\rm \bf C})$ Haar measure, and an orthonormal basis for the resulting Hilbert
space of $L^2$ functions is given by the simple spin networks:
\be
s^{\{\rho_e\}}(g_e,x_v)=
\prod_{e}K_{\rho_e}(x_{s(e)},g_e.x_{t(e)})=
\prod_{e}\langle\rho_ex_{s(e)}(j=0)|g_e|\rho_ex_{t(e)}(j=0)\rangle,
\label{simplespinnet}
\ee
where $\mid\rho x(j=0)\rangle$ is the vector of the $\rho$ representation
invariant under the $SU(2)$ subgroup leaving the vector $x$
invariant. This same Hilbert space for kinematical states
comes out of a canonical analysis in an explicitly covariant
framework\cite{link}. Now the main issue is: what kind of transition
amplitudes are those defined by the Barrett-Crane
model? 
We argue that it is a realization of the
projector operator for quantum gravity\cite{causal}, as argued also in \cite{projector}\cite{ansdorf}.
The task is to locate clearly where in the model the ${\rm \bf Z}_2$
symmetry relating positive and negative proper times is implemented.
As we said, this symmetry originates from
the symmetry $B\rightarrow -B$ in the bivector field,
and in our discretized context corresponds to a
change in the orientation of the triangles in the simplicial
manifold, and consequently of the
tetrahedra. 
Let us use the unique decomposition of $SL(2,{\rm \bf C})$
representation functions of the 1st kind $K$ into representation
functions of the 2nd kind $K^{\pm}$ and write the $K$ functions as
 \bes
K^{\rho_{ij}}(x_i,x_j)=\frac{2\sin(\eta_{ij}\rho_{ij}/2)}{\rho_{ij}\sinh\eta_{ij}}
=\frac{e^{i\,\eta_{ij}\,\rho_{ij}/2}}{i\rho_{ij}\sinh\eta_{ij}}\,-
\,\frac{e^{-\,i\,\eta_{ij}\,\rho_{ij}/2}}{i\rho_{ij}\sinh\eta_{ij}}\,=\,K^{\rho_{ij}}_{+}(x_i,x_j)+K^{\rho_{ij}}_{-}(x_i,x_j)=
\nonumber \\ =
K^{\rho_{ij}}_{+}(x_i,x_j)+K^{\rho_{ij}}_{+}(-x_i,-x_j)\,=\,K^{\rho_{ij}}_{+}(x_i,x_j)+K^{\rho_{ij}}_{+}(x_j,x_i)=
K^{\rho_{ij}}_{+}(\eta_{ij})+K^{\rho_{ij}}_{+}(-\eta_{ij})\,\,\,\,
\label{eq:Ksplit} \ees and make the following alternative expressions
of the same $Z_2$ symmetry manifest: \be
K^{\rho_{ij}}(x_i,x_j)=K^{\rho_{ij}}(\eta_{ij})=
K^{\rho_{ij}}(-\eta_{ij})=K^{\rho_{ij}}(-x_i,-x_j)=
K^{\rho_{ij}}(x_j,x_i)=K^{-\rho_{ij}}(x_i,x_j). \ee We see that the symmetry
characterizing the projector operator is indeed implemented as a
symmetry 1) under the exchange of the arguments of the $K$ functions,
so the model does not register any ordering among the
tetrahedra in each 4-simplex, 2) under the change
of orientation of the two tetrahedra sharing the
triangle, so it does not register its orientation either, 3) under a change in sign of the distance between
the two points on $H^+$, i.e. in the way the model uses
upper and lower hyperboloids ($\eta\in H^+ \leftrightarrow
-\eta\in H^{-}$), 4) under
the exchange of a representation $(0,\rho)$ and its dual $(0,-\rho)$.
\section{Implementing causality: a causal transition amplitude}
We turn now to the problem of implementing causality in the
Barrett-Crane model, i.e. to break consistently the identified
$Z_2$ symmetry. This consistent restriction is
found by requiring that
the resulting amplitude has stationary points corresponding to
good simplicial Lorentzian geometries\cite{causal}. 
For a given triangulation $\Delta$, the amplitude reads
\be
A(\Delta)=\sum_{\epsilon_t=\pm1}\int\prod_t\rho_t^2\textrm{d}\rho_t
\prod_T A_{e}(\{\rho_t,t\in T\})
\prod_s\int_{({\cal H}^+)^4} \prod_{T\in s}\textrm{d}x^{(s)}_T
\left(\prod_{t\in s}
\frac{\epsilon_t}{i\rho_t\sinh\eta_t}\right)
e^{i\sum_{t\in s}\epsilon_t\rho_t\eta_t}. \label{eq:am}
\ee
The action for a single (decoupled) 4-simplex is then \be
S=\sum_{t\in s}\epsilon_t\rho_t\eta_t=\sum_{t=(ij)\in s}\epsilon_{ij}\alpha_i\alpha_j\rho_{ij}\theta_{ij}, \ee with the angles
$\theta_t$ constrained by the
Schl{\"a}fli identity, that can be enforced by a Lagrange multiplier $\mu\in{\rm
\bf R}$, and its stationary points are defined by \be 
\epsilon_{ij}\alpha_i\alpha_j\,=\,sign(\mu)\;\;\;\;\;\;\;\;
\rho_{ij}\,=\,|\mu|\,A_{ij}.
\label{localorient}
\ee
This means that the area of the triangles are given (up to scale) by
the representation labels $\rho_{ij}$ and that we have a consistency
relation between the orientation of the tetrahedra $\alpha_i$, the orientation
of the triangles $\epsilon_{ij}$ and the global orientation
of the 4-simplex $sign(\mu)$.
Now we can extend this orientation to the whole spin foam, imposing
that if a tetrahedron is past-oriented for one 4-simplex,
then it ought to be future-oriented for the other sharing it.
A consistent orientation is thus a choice of $\mu_v$ and $\alpha_{T,v}$ (for each tetrahedron $T$ attached to a
4-simplex $v$) such that:
$\forall T,\,\mu_{v_1}\alpha_{T,v_1}=-\mu_{v_2}\alpha_{T,v_2}$
where $v_1$ and $v_2$ are the two 4-simplices sharing $T$.
This is also equivalent to requiring an oriented dual 2-complex. 
Now we can fix the variables $\epsilon_t$, and thus break the
$Z_2$ symmetry that erases causality from the model (average over all possible orientations
of the triangles), to
the values corresponding to the stationary points in the a-causal
amplitude (of course we do not impose the
equations of motion for $A_t$ and $\theta_t$). This leads to a {\it causal
amplitude}\cite{causal}, constructed by picking from \ref{eq:am} only the terms corresponding to the chosen
$\epsilon_t$:

\be
A_{causal}(\Delta)
=\prod_s\int_{({\cal H}_+)^4} \prod_{T\in s}\textrm{d}x^{(s)}_T
\prod_{t\in s}
\frac{\epsilon_t}{i\rho_t\sinh\eta_t}\int\prod_t\rho_t^2\textrm{d}\rho_t
\prod_T A^T_{e}(\{\rho_t\}_{t\in T})\,e^{i\,\sum_t\,\rho_t\,\sum_{s|t\in
s}\theta_t(s)},
\label{orientampli}
\ee
with suitable boundary terms, to be understood within a sum over
oriented 2-complexes or triangulations, provided with a consistent causal structure
$(\Delta,\{\mu_v,\alpha_T,\epsilon_t\})$.
The causal partition function is then basically of the form:
\be
Z_{causal}\,=\,\sum_\Delta\,\lambda(\Delta)\,
\int{\mathcal{D}}\theta_t(\Delta)\,
\int{\mathcal{D}}A_t(\Delta)\,e^{i\,S_{R}^\Delta(A_t,\theta_t)}
\ee
What are its features? 1) It is a simplicial realization of the sum-over-geometries approach
to quantum gravity, for a first order Lorentzian
Regge action, with areas and dihedral angles being independent
variables, and with a precise assignment of a measure and an additional
sum over causally well-behaved triangulations, where the causal relations are encoded in the
orientation;  2) the geometric variables have a natural algebraic characterization in the representation theory of the Lorentz group, and the
combinatorial data used are from the 2-complex dual to the
triangulation only, so the model is a spin foam model;
3) it realises the general definition given for a causal spin foam model (it is, to the best of our
knowledge, its first non-trivial example), except for the use of
the full Lorentz group instead of its $SU(2)$ subgroup; 4) it
identifies causal sets (given by the ``1st layer'' of the dual
2-complexes) as the fundamental discrete structures on which
quantum gravity has to be based, as in
\cite{Sork}, but it contains additional metric
data intended to determine a consistent length scale,
which, in the traditional causal set approach, is meant to be
obtained by ``counting only''; this is a
particular case included in the model and obtained by fixing all the geometric data to some arbitrary
value; 5)
also, if we fix all the geometric data to be those obtained from
a fixed edge length, then what we obtain is the conventional sum
over triangulations in the dynamical triangulations approach to
quantum gravity, for Lorentzian triangulations; the
additional integrals, if restored, can be
intepreted as providing a sum over proper times, usually not
implemented in that approach.
In the new formulation presented above, the Barrett-Crane model fits \cite{causal}
into the general scheme of quantum causal sets (or quantum causal
histories)\cite{fot2}, being its first explicit non-trivial example. 
Consider an oriented graph, restricted to be 5-valent and to not include closed cycles of
arrows, identified with the first {\it stratum} of the spin foam 2-complex. Interpreting the arrows as representing causal relations, this
is a causal set, the orientation of the links
reflecting the ordering relation among the vertices.
Because of the restriction on the valence, it can be decomposed into
building blocks, since for each
vertex only one out of four possibilities may be realised,
corresponding to the $4$ possible Pachner moves ($4-1$, $3-2$, and
their reciprocal), in the dual simplicial interpretation, giving the evolution of a 3-dimensional simplicial
manifold in time.
The crucial point is the identification of the direction of
the arrow in the causal set with the orientation of the tetrahedron it
refers to.
The quantization is then the assignment of the
Hilbert spaces of intertwiners among four given simple continuous representations of the Lorentz
group, representing the possible states of the tetrahedra in the
manifold, to the {\it arrows} of the causal set and of the causal Barrett-Crane amplitude defined
above to its nodes, as evolution
operator. Hilbert spaces that are
a-causal to each other can be tensored together, so, in particular,
for two a-causally related tetrahedra we can tensor the corresponding
intertwiners, so obtaining open spin networks with more than one
vertex for each a-causal set of events. In
each of the building blocks both the source and target arrows form an
a-causal set\cite{fot2}, and together form a
complete pair, so they are in turn linked by a causal relation in
the poset of a-causal sets defined on the edge-poset\cite{fot2}; to each causal relation among complete pairs, i.e. to each of the
building blocks of the edge-poset, we associate the causal
Barrett-Crane amplitude for a single 4-simplex. Also operators
referring to a-causal sets that are not causally
related to each other can be tensored together, so
composite states constructed evolve according to
composite evolution operators built up from the fundamental ones.
We have already stressed from the beginning that a sum over
2-complexes is necessary to restore the full dynamical
content of the gravitational theory. In this causal set picture this means that we have to
construct a sum over causal sets interpolating between given boundary
a-causal sets $\alpha$ and $\beta$, each poset weighted by the causal Barrett-Crane amplitude, so that the full
evolution operator is
${\mathcal{E}}_{\alpha\beta}=\sum_c\lambda_c A^{c}_{\alpha\beta}$. 
 The properties of ${\mathcal{E}}_{\alpha\beta}$ and $A^{c}_{\alpha\beta}$ (antisymmetry, reflexivity, transitivity, unitarity)
are discussed in \cite{causal}.
\section{Conclusions}
We described briefly the Lorentzian Barrett-Crane model, which is the
most studied spin foam model for 4-d quantum gravity, and the classical and quantum description of spacetime geometry behind its
formulation\cite{causal}.  We explained why it has to be considered as providing the
physical inner product between quantum gravity states, being a
covariant 
realization of the
projector operator onto physical states, and we identified
explicitly in the model the $Z_2$ symmetry that characterizes it.
We have shown how to break this symmetry consistently to
obtain a spin foam realization of a
causal transition amplitudes between quantum gravity states. The
resulting spin
 foam model\cite{causal} is a path integral for Lorentzian 1st order Regge
calculus with an algebraic characterization of all the geometric variables and with a clear definition of the integration
measure. It is the first non-trivial example of
a causal spin foam model, fits into the framework of quantum
causal histories and links several areas of
research: canonical and
sum-over-histories formulation of quantum gravity, Regge calculus,
causal sets and dynamical triangulations.

\end{document}